# The WSO, a world-class observatory for the ultraviolet

M.A. Barstow[*a], L. Binette[b], N. Brosch[c], F-Z. Cheng[d], M. Dennefeld[e], A.I. Gomez de Castro[f], H. Haubold[g], K.A. van der Hucht[h], N. Kappelmann[i], P. Martinez[j], A. Moisheev[k], I. Pagano[l], E.N. Ribak[m], J. Sahade[n], B. Shustov[o], J.-E. Solheim[p], W. Wamsteker[q], K. Werner[i], H. Becker-Ross[r] and S, Florek[r]

[a]University of Leicester; [b]Tel Aviv University; [c]UNAM Mexico, [d]CfA-USTC, Hefei;
[e]Institute d'Astrophysique, Paris; [f]IAG, Madrid; [g]UN-OSD Vienna, [h]SRON, Utrecht;
[i]Universität Tübingen; [j]SAAO, [k]Lavochkin Association, [l]Catania; [m]Technion, Haifa,
[n]La Plata; [o]INASAN, Moscow; [p]University of Trömso, [q]ESA, Madrid,
[r]Inst. Spectrochemistry and Applied Spectroscopy, Berlin.

## ABSTRACT

The World Space Observatory is an unconventional space project proceeding via distributed studies. The present design, verified for feasibilty, consists of a 1.7-meter telescope operating at the second Largangian point of the Earth-Sun system. The focal plane instruments consist of three UV spectrometers covering the spectral band from Lyman alpha to the atmospheric cutoff with R~55,000 and offering long-slit capability over the same band with R~1,000. In addition, a number of UV and optical imagers view adjacent fields to that sampled by the spectrometers. Their performance compares well with that of HST/ACS and the spectral capabilities of WSO rival those of HST/COS.

The WSO, as presently conceived, will be constructed and operated with the same distributed philosophy. This will allow as many groups and countries to participate, each contributing as much as feasible but allowing multi-national participation. Although designed originally with a conservative approach, the WSO embodies some innovative ideas and will allow a world-class mission to be realized with a moderate budget.

## 1. INTRODUCTION

The considerable scientific success of the IUE observatory and successor instruments such as the GHRS and STIS spectrographs on-board HST amply demonstrate the importance of the far ultraviolet wavelength range, from ~100 to 300nm, to modern astronomy. Of particular importance has been access to high resolution R~40,000-100,000 echelle spectra providing an ability to study the dynamics of hot plasma and separate multiple stellar and interstellar absorption components. However, there are a few emerging problems with this provision. First, since HST has UV, optical and infra-red capabilities, the available observing time is divided across these bands. In addition, the total observing time is oversubscribed by a factor ~10. Hence demand for access to UV observations far outstrips what is available from HST. Secondly, although it is currently planned to operate HST until 2010, the STIS instrument is already operating on the redundant electronic systems and the instrument will not be serviced. Therefore, it is necessary to plan for a replacement of this high resolution UV capability.

The World Space Observatory is the proposed solution to the problem of future access to high-resolution far-UV spectroscopy. The planned instrument sensitivity will exceed that of HST/STIS by a factor 5-10, but all the observing time will be available for UV astronomy. Furthermore, it is planned to operate the mission at the L2 point of the Earth-Sun system providing a significant increase in operational efficiency over low Earth orbit. Taking all these factors into account will yield a net increase in UV productivity of a factor ~40-50, compared to HST/STIS. The present mission design consists of a 1.7-meter telescope. The focal plane instruments consist of three UV spectrometers covering the spectral band from Lyman alpha to the atmospheric cutoff with R~55,000 and offering long-slit capability over the same band with R~1,000. In addition, a number of UV and optical imagers view adjacent fields to that sampled by the spectrometers. The imaging performance compares well with that of

HST/ACS while the spectral capabilities are comparable to the HST/STIS echelle modes. However, with a smaller number of instruments in the focal plane, compared to HST, the required number of optical elements in each subsystem is reduced. Hence, the WSO delivers considerably enhanced effective area. This thoughput is similar to that offered by the Cosmic Origins Spectrograph (COS) at the longer wavelengths, which is to be mounted on HST in 2004, but COS has an inferior spectral resolution (R~24,000 max). With a 2006/7 launch date, WSO is ideally placed to provide follow-up studies of the large number of UV sources expected from the GALEX sky survey.

The WSO has been conceived as an unconventional space project, developed via distributed studies rather than under the auspices of a single agency. It will be constructed and operated with the same philosophy, allowing many groups and countries to participate, each contributing as much as feasible from their national resources. Although designed with a conservative approach, to keep mission costs within a moderate budget, the WSO embodies some innovative ideas that will allow a world-class mission to be realized.

## 2. THE WSO MISSION CONCEPT

There are several driving principles behind the design of the WSO mission. The plan is to operate a 1-2 meter class telescope in Earth orbit with a spectroscopic and imaging capability specific to the UV wavelength range. The telescope should have a high throughput and the mission optimized for operational and orbital efficiency. However, the cost of construction, launch and operation should be minimized without affecting the scientific excellence of the mission products. Consequently, the degree of technological development should be limited and use of existing technology be employed where possible. WSO will provide access to a front-line facility for basic space science for the wider international astrophysics and planetary science communities. We discuss below the various mission subsystems and their degree of technological development.

**2.1 The T170-M telescope**

The heritage for the WSO telescope design[1] is the Russian-led international space observatory ASTRON, launched in 1983. It functioned for 6 years and was the first UV telescope placed into a highly eccentric orbit. This led to the proposed development of the Spectrum UV mission with Russian, Ukrainian, German and Italian partnership. However, the major social and economic changes taking place in the former Soviet Union countries has reduced funding of the Russian space programme and limited progress in all space projects. The experience gained in the development of the Spectrum UV T170 telescope is, therefore, now being applied to a modified version (the T170M) for the much broader international scope of the WSO.

The structure of the T170M telescope is shown in Fig. 1, with the principle structural elements; the Primary Mirror Unit (PMU), the Secondary Mirror Unit (SMU) and the instrument compartment. The SMU is attached to the telescope structure with a spider. The optical design is a Ritchey-Chretien type with a 1.7m hyperbolic primary mirror and engineering model of this unit for the T170 design (with a spherical Sitall mirror) has successfully passed through a number of environmental tests. Modifications for the WSO centre on reducing the overall mass of the system below 1650kg. Consequences of this are a shorter relative focal length, although the f number (10) remains the same, and reduced distances between the mirrors and between the primary and optimum focal surface (Table. 1).

|  | T170 | T170M |
|---|---|---|
| Aperture diameter | 1.7m | 1.7m |
| Telescope focal length | 17.0m | 17.0m |
| $R_1$ (primary radius) | -9.333m | -7.820m |
| $\varepsilon_1$ (eccentricity) | 1.050237 | 1.029508 |
| SM light passing diameter | 458mm | 399.1mm |
| $R_2$ (secondary radius) | -3.216m | -2.214m |
| $\varepsilon_2$ (eccentricity) | 3.612442 | 2.848076 |
| Space between mirrors | -3.5m | -3.057m |
| Obscuration with optimum baffles | 0.347 | 0.316 |
| Space between PM apex and optimum focal surface | 0.75m | 0.65m |

Table 1. Summary of the technical details of the T170M telescope design and comparison with the earlier T170.

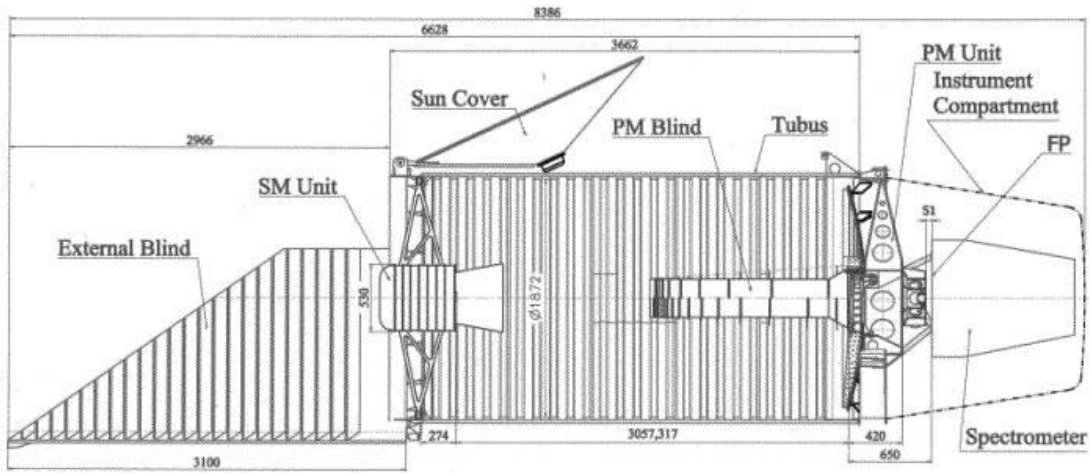

Fig. 1: Schematic diagram of the WSO 1.7-m aperture UV telescope.

The secondary mirror is design includes capabilities for in-flight alignment and focusing. These are controlled by means of: 1) a 1mm displacement of the secondary along two mutually perpendicular axes (in a plane normal to the optical axis) in steps of 0.01mm; 2) focusing with axial movements of the secondary mirror within ±5mm from the normal position, in steps of 0.002mm. Both primary and secondary mirrors will be coated with Al and a $MgF_2$ layer, to yield an interference maximum at 115nm.

The primary mirror is supported by a central mount and includes a system of heaters on its lower side to minimize the thermal gradients along the mirror during operations. A support structure consisting of 12 profiles Invar bars and is covered by corrugated Al sheets to reduce thermal distortions of the tube. An rms wavefront error of less than 13nm is required of the system to deliver diffraction limited optical performance. Figs 2. and 3. show the predicted telescope performance as a function of distance from the centre of the field. The diffraction limit corresponds to 0.035 arcsec.

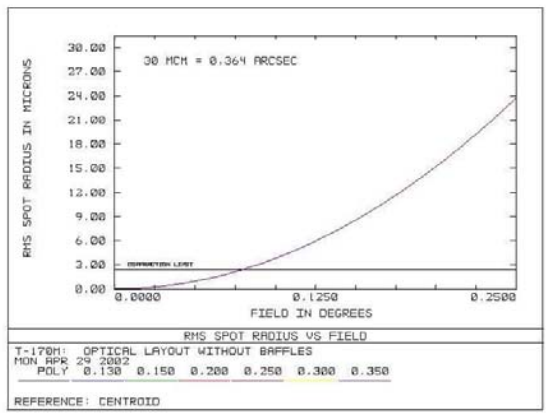
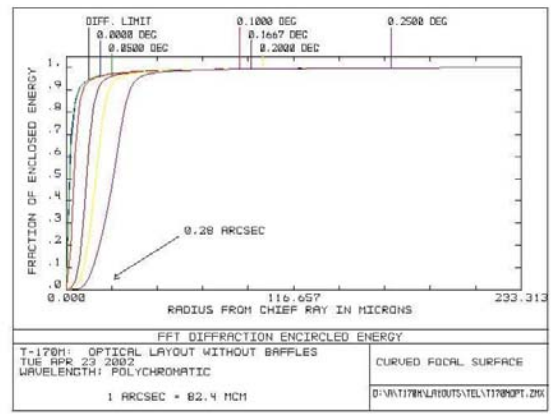

Fig. 2: Predicted rms spot radius as a function of off-axis position for the T170M design.

Fig. 3: Fraction of enclosed energy at various off-axis positions in the T170M telescope field.

## 2.2 The high resolution double echelle spectrograph (HIRDES)

The UV spectrometer[2] (Fig. 4) comprises three different single spectrometers. Two of these are echelle instruments, designed to deliver high spectral resolution, and the third is a low dispersion long slit instrument (LSS). At high dispersion, the 110 to 320nm waveband of the WSO will be divided into two, the UV (UVES, 178-320nm) and VUV (VUVES, 103-180nm). The fundamental concept of HIRDES is based on the design heritage of the ORFEUS missions (Orbiting and Retrievable Far and Extreme Ultraviolet Spectrometer, mounted on the ASTRO-SPAS free flyer), successfully flown on two STS flights in 1993 and 1996.

Each of the three spectrometers has its own entrance slit, lying in the focal plane of the T-170M telescope on a circle with diameter 100 mm. The three optical trains are not used for simultaneous observation, but in sequential mode. This is managed by satellite motion with a pointing stability requirement of 0.1 arcsec to be monitored by three Fine Guidance Sensors. Each of the three sub-instruments includes optical elements to form the spectral imaging and uses a main and a redundant detector (baseline: MCP/WSA detector) for observation. The main and the redundant detectors are placed closely together in a L-shaped detector housing. A servo-driven mirror in front of the detectors will illuminate either of the detector apertures. Between this servo mirror and the redundant detector an additional servo-driven gray filter will be fitted, to reduce the intensity, so the redundant detector can be used to observe brighter objects if the front-end electronics of the main detector is saturated. Technical details of the spectrographs are listed in tables 2, 3 and 4, while the optical layout is in Fig. 4.

| Entrance aperture | circular 80 μm |
|---|---|
| Collimator mirror | toroidal, R1 = 1608 mm, R2 = 1593 mm, circular 80 mm |
| Echelle grating | 40 grooves/mm, 66.9° blaze, 90 mm x 190 mm |
| Cross dispersion prism | 12° quartz, double pass, 100 mm x 110 mm |
| Camera mirror | spherical, R=1600 mm, 110 mm x 130 mm |
| Detector: | 30 mm (echelle dispersion), 30 μm resolution (pixel-width) |
|  | 40 mm (prism dispersion), 40 μm resolution (pixel-height) |
| Resolving power | R = 50000 (3 pixel criterion) |

Table. 2: Technical details of the UV echelle spectrograph (UVES)

| Entrance aperture | circular 80 μm |
|---|---|
| Collimator mirror | parabolic, R = 1600 mm, 6°-off-axis, circular 80 mm |
| Echelle grating | 65 grooves/mm, 71° blaze, 90 mm x 270 mm |
| Wadsworth grating | toroidal, 625 grooves/mm, R1 = 1698 mm, R2 = 1695 mm, 90 mm x 120 mm |
| Detector: | 30 mm (echelle dispersion), 30 μm resolution (pixel-width) |
|  | 40 mm (Wadsworth dispersion), 40 μm resolution (pixel-height) |
| Resolving power | R = 55000 (3 pixel criterion) |

Table. 3: Technical details of the VUV echelle spectrograph (VUVES)

| Entrance aperture | rectangular, 80 μm x 6 mm |
|---|---|
| Rowland grating | aberration corrected concave grating, R ≈ 800 mm |
| Detector: | 80 mm (grating dispersion), 15 μm resolution (pixel-width) |
| Resolving power | examples: R = 1000 (3 pixel criterion), 110 nm to 350 nm |

Table. 4: Technical details of the long slit spectrograph (LSS)

The original spectrometer design for ORFEUS incorporated a Z-stack MCP detector with a wedge-and-strip anode readout. The leading two MCPs are standard resistivity plates with 12.5μm pores, 15° bias angles (for HIRDES) and L/D ratio of 80:1, while the rear MCP is a low resistance L/D 40:1 unit. The useful detector area is 40mm x 30mm. The spatial position of the impinging photons corresponds to the focal point of the generated electric charges at the third MCP outlet and is detected by means of the wedge-and-strip anode. From four electrodes the charges of an electron cloud are measured with ADCs with an oversampling technique. From these four charge values the exact coordinates of the centre of the electron cloud are calculated in the ICU. The spatial resolution in the WSA plane depends on the ADC resolution. With 12 bit ADCs a minimum resolution of 30 μm (x-direction) and 40 μm (y-direction) can be achieved, corresponding to 1024 x 1024 pixels in the detection plane. With 14 bit ADCs a minimum resolution of 7 μm (x-direction) and 10 μm (y-direction) can be achieved, corresponding to 4096 x 4096 pixels. These resolutions are well matched to the spectrometer design requirements.

Adoption of a common detector design for the long slit dispersion spectrometer poses a surprising problem. Although the instrument operates at lower spectral resolution, the full 110-350nm spectral range cannot be accommodated on a single detector of the above dimensions. Hence, there is a potential trade-off between the desired resolving power (R~1000) and spectral coverage. One possible solution is to use the higher spatial resolution detector design, to allow the full wavelength range to be compressed into a smaller detector field. We propose to adopt an MCP detector design, using a Vernier anode readout[3], which has recently been proven in-flight on board the J-PEX high resolution EUV spectrometer[4,5].

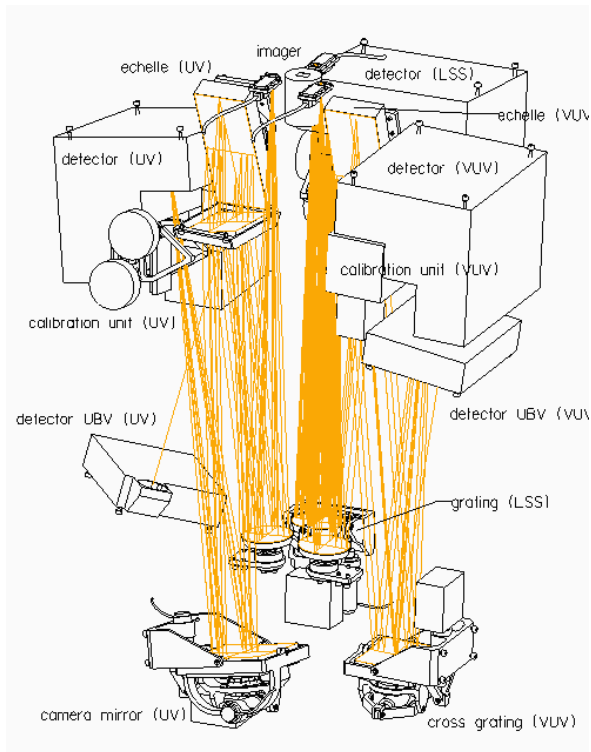

Fig.4: Optical layout of the 3 HIRDES spectrometers.

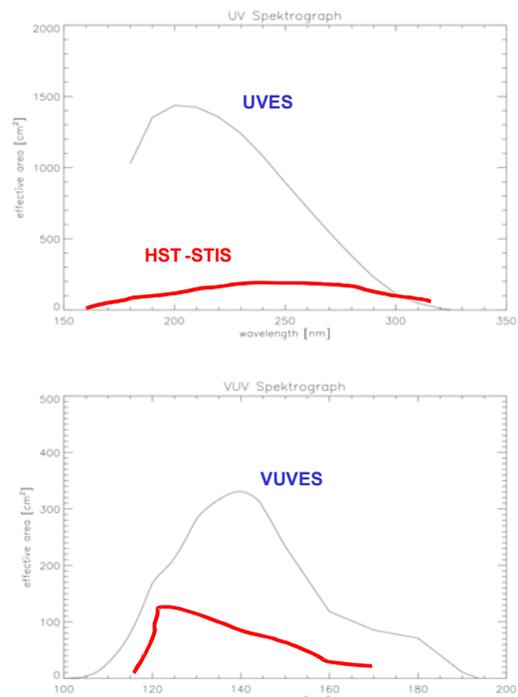

Fig. 5: Predicted effective areas of the UVES (top) and VUVES (bottom) spectrographs, compared with that of HST/STIS.

## 2.3 WSO performance with the HIRDES spectrographs

To be a viable replacement for HST, the overall capabilities of the WSO prime spectroscopic instruments need to compete with the performance of the STIS and COS spectrographs, which they will eventually supercede. In having a mission dedicated to a narrower waveband than HST, which also has coverage extending into the visible and infrared, there are several advantages. First, a simpler optical layout of the spectrograph systems is available. A resulting reduction in the number of necessary reflections, to accommodate the instruments in the available space, leads to lower light losses. Second, **all** the available observing time can be devoted to UV astronomy. Third, the intention of placing the telescope at the L2 Sun-Earth point yields a more efficient duty cycle, compared to low Earth orbit. Fig. 5 shows the comparison in the effective area of HST with the STIS spectrograph and the WSO UVES and VUVES instruments, while Table 5 compares the WSO performance with that COS.

WSO has an effective area ~4-10 times that of HST/STIS, while providing a resolving power approaching that of the highest STIS echelle capability. When this gain is folded with the improvements in available observing time, the overall productivity will be a factor 40-50 higher than STIS. In contrast, at the shorter wavelengths the effective area of COS is larger than VUVES (although again the observing time factors and the larger wavelength coverage make the productivity comparable) but falls below the performance of UVES at the longer wavelengths. However, the COS resolving power is a factor of two lower.

| λ (nm) | λ/δλ | $A_{eff}$ (cm$^2$) | λ (nm) | λ/δλ | $A_{eff}$ (cm$^2$) |
|---|---|---|---|---|---|
| 115-150 | 24,000 | 2,200 | 102-180 | 55,000 | 350 |
| 140-178 | 24,000 | 1,200 | | | |
| 175-200 | 24,000 | 600 | 178-310 | 50,000 | 1450 |
| 200-250 | 24,000 | 650 | | | |
| 250-320 | 24,000 | 450 | | | |

Table. 5: Comparison of the predicted performance of the WSO UVES and VUVES spectrographs (right) with the HST COS instrument (left).

### 2.4 WSO focal plane imagers

Although the primary science of the WSO mission is spectroscopic, there is an important role for high spatial resolution UV imaging of the sky. Therefore, it is planned to include a complement of UV imaging detectors[6] in the focal plane, to provide serendipitous science during spectroscopic observations as well as planned studies of specific target areas. Much of the available volume in the focal plane, immediately behind the primary mirror, is occupied by the HIRDES. This leaves only a very narrow space (10cm diameter cylinder) on the telescope axis that can be used for a direct imager, which samples the best diffraction limited resolution of the optical system. This detector (the central field camera, CFC) will require a fixed filter, since there is no room for a filter wheel.

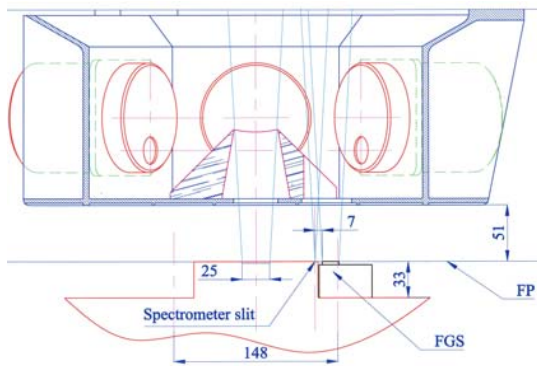

Fig. 6: Side view of the optical bench, showing three UV imagers, the redirecting prism and one possible arrangement for the fine guidance sensor.

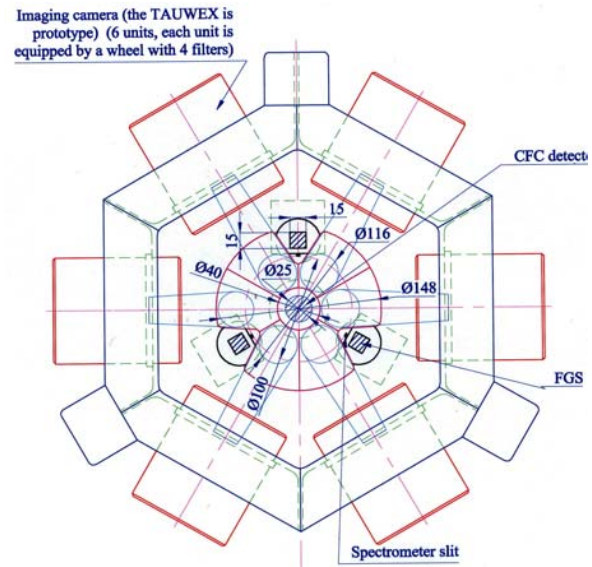

|  | IFOV(arcsec) | FOV(arcsec) |
|---|---|---|
| CFC | 0.03 (2.5μm) | 60.0 |
| HSI | 0.15 (12.5μm) | 300.0 |
| FGS | 0.054 | 55.2 |

Table. 6: Performance of UV and optical imagers.

Fig. 7: Proposed focal plane optical bench, accommodating 6 UV imagers in addition to the central field camera. The spectrograph slit positions are also shown

One solution to the accommodation problem for additional imagers is to redirect the beam through 90° by a single reflection, as shown in figs. 6 and 7. An option currently under investigation is to operate 3 imagers at f/10, each with a flat mirror pickup, and 3 at f/50, using magnifying relay of two mirrors. Thus the f/50 cameras would operate at optimal spatial sampling and there would be more space in the aperture to accommodate the fine guidance sensors. These HSI cameras are based on the TAUVEX design for Spectrum X-gamma and each is equipped with a

4-element filter wheel. Combining two wheels with 3 filters plus one open position will provide 6 colours plus a completely open position. As can be seen from table 6, which summarises the target performance for the imagers (for a plate scale of 12.06 arcsec mm$^{-1}$), the demands on detector spatial resolution are stringent. The J-PEX detector design, discussed above for the LSS, provides the best available but is ultimately limited by the MCP pore size of 6μm.

## 3. WSO PROJECT ORGANISATION

The WSO concept was originally proposed at the 7th UN/ESA Workshop for Basic Space Science[11] and formulated in the context of the needs of Basic Space Science in the developing countries1. It was defined as a mutually beneficial and valid scientific mission for all scientists in the world and not confined only to the developing world. In the context of the Long Range Planning of the European Space Agency, an Assessment Study[7,8] was carried out, evaluating the mission for the ultraviolet telescope described in section 2.

The WSO/UV will be an Astrophysical Observatory rather than a targeted mission for a single science goal. Its optimum launch date is in the 2006/7 time frame to coincide with the near end of HST operations and the large demand for new UV spectroscopic and imaging observations that will inevitable be stimulated by the GALEX UV sky survey. Consequently, it fills an important gap in the capabilities available to the astronomical community[9,10]. The sensitivity and resolution supplied by the mission are matched to the capabilities of the current generation of X-Ray Observatories in space, the 8m size ground-based telescopes, and future IR missions (e.g. NGST). WSO will not duplicate any capability available from the ground nor in space. The operations philosophy of the WSO/UV [3] will assure that it can support a very dynamic science program over the extended life span of up to 10 years. This is essential for a mission of this nature, which permits full participation from its very beginning of the scientific and technical community on a worldwide scale.

The WSO project is not led by a single agency but is being implemented as a distributed project with participation and contributions from a larger number of countries than is usual in space research, although it hoped to have space agency contributions to deal with certain quality control and management issues that may be crucial to mission success. Participation in the project is organized through a network of National WSO Working Groups, the chair of each of these acting as a national project representative on the WSO Implementation Committee (WIC). The author list for this paper comprises the current (but not necessarily final) membership of the WIC, plus several key instrument scientists. The work included in this paper is partly the result of several studies and meetings organized under the auspices of the WIC. Key milestones include:

- *May 2000*: ESA Assessment Study completed (**CDF-05**)
- *August 2000*: **IAU**: WSO incorporated in Future Large Scale Facilities Working Group activities.
- *December 2000*: **DLR**: Phase A study started to evaluate the adoption of HIRDES spectrograph to WSO/UV (including technology cut-off 2002)
- *December 2000*: **JPL**: Team X Red Team Review of WSO/UV completed. Confirms technical feasibility and scientific relevance.
- *April 2001*: **ASI**: Science Interest Definition meeting
- *April 2001*: **DLR** : HIRDES phase A review meeting
- *May 2001*: **China**: Xiang-Shan Workshop on WSO/UV (9 -11 May) in Beijing
- *May 2002:* **Russia**: Phase A study initiation meeting, Moscow (19-23 May)

The phase A study of the WSO mission, including an evaluation of the science goals, is now underway and is planned to end on November 30$^{th}$ 2002.

The novel nature of the organization of the WSO as a project is of great importance for the participation of less developed nations who may not yet have participated in an active space programme. In many developing countries, the significant investment in education is not bearing the desired fruits. The reason for this is closely associated with the fact that participation in advanced science can only function efficiently in the industrialised world. Consequently, investment in education often results only in the creation of a consumer market, without the creation of the professionally well-formed, culturally and intellectually identifiable, and academically oriented cadre of scientists that is necessary for sustainable development. It is clear that there exists a strong need for a fruitful interplay between the academic and the commercial sector of the population, not solely driven by the last. Without the relatively small fraction of the population driven by the pursuit of intellectual progress, such a synergetic process cannot be sustained in the context of its original socio-cultural climate. For the required accelerated and sustainable development, essential for all future projections of the world economy to have any chance of success, quantum leaps in development in various areas are essential. Space sciences supply a unique medium to accomplish such forward jumps.

WSO/UV embodies a three-fold goal. First, to create opportunities for participation at the frontiers of Basic Space Science, on a sustainable basis at the national level, by all countries in the world, without the need for excessive investment. The observatory will stimulate the creation of an academically mature and competitive cadre within 5 to 10 years after inception by offering equal opportunities to astronomers all over the world. Second, to generate the possibility that engineers/specialists from the developing countries to work with the engineering groups for short periods of time and learn to built, handle and test etc. the WSO/UV hardware in an academic environment. Third, to support worldwide collaboration and ensure that the study of the Universe from space can be maintained in a sustainable way by scientists from all countries. This will make a reality in the scientific world of the visionary principle that *space is the province of all mankind*.

The WSO/UV will help to drive internal development in countries and will contribute significantly to infrastructure and higher education benefits to both the political and general population in the developing countries. As the WSO/UV is also of great interest for the scientists in the developed world it could generate a very strong stimulus to enhanced and sustained development, through a strong mutually beneficial, interplay between the scientific communities. This will not only have its impact in the sciences but equally, or even more so, for politics, the humanities and applied technologies, because it will create locally identifiable familiarity with the Basic Space Sciences.

## 4. WSO SCIENCE GOALS

This paper is intended to be mainly a technical description of the WSO mission and project. However, given its nature as an observatory class facility, that will be available to any scientist worldwide, it is useful to consider some of the scientific contributions it is expected to make beyond HST.

### 4.1 Planetary system science

The unique aspects of high-spectral resolution and excellent imaging capabilities of WSO will allow unique research opportunities in the Solar System. Some of these projects will include studies of the gaseous atmospheres of the giant planets (spectroscopic diagnosis of individual features, e.g., at the edge of Jupiter's Great Red Spot by using the small apertures of HIRDES; longitudinal distribution of species using the long-slit; monitoring of auroral activity using the UV imagers). Others will concentrate on comets, following similar studies done with IUE and HST. A new topic will probably be diagnosing the ices in the outer Solar System; these studies will search for signatures of molecular species trapped in the surface ices through low-resolution spectroscopy. Possible targets will be Europa and Ganymede, Miranda, Pluto and Charon, and Kuiper Belt objects. However, these are just examples and other outer Solar System targets are equally likely to be proposed. It is expected that this way, one could get a handle on the photo-evolution of ices, on material implantation onto icy satellites, and on the creation of organic materials in the outer Solar System.

### 4.2 Stellar science

The complete life cycle of stars can be studied with many new discoveries to be expected. Not only the outer photospheric, chromospheric and coronal domains of a star can be studied, but also high frequency phenomena which will give information on the interiors of stars can be studied very effectively. One of the fields in which the project will represent a major breakthrough is in the modifications to stellar evolution as a consequence of the multiplicity in star systems, such as the effects of close binary mass exchange and accretion on condensed objects. Also many aspects of the rapidly changing shock phenomena in Young Stellar Objects and the physical mechanisms driving jets in such objects, form an extremely exciting area of application of the WSO/UV.

### 4.3 White dwarfs and the local interstellar medium

The high dispersion echelle mode available with WSO represents an essential tool for pursuing a number of important astrophysical investigations. Chief among these are the possibility of variations in the local D/H ratio, the structure, composition and ionization of the LISM, and finally, a host of fundamental physical processes in the atmospheres of white dwarfs. All three of these topics are best studied in the UV, at high dispersion, using the smooth well-understood continuum fluxes of nearby hot white dwarfs. Important goals will be to: measure the D/H ratio within the local interstellar medium (LISM) and define its variation; Map the 3D structure of the LISM and determine its composition and ionization state; define the evolutionary history of the hot white dwarf stars through detailed modeling of their photospheric composition and structure; study the pattern of circumstellar material surrounding the white dwarfs and their interaction with the ISM.

### 4.4 Star formation

Outflow is ubiquitous during star formation: it is observed from the very early phases (the so-called Class 0 sources) to the late stages of the pre-main sequence evolution. Bipolar flows are widely assumed to be powered by the gravitational energy released during the accretion process and the magnetic field is often hypothesized as the physical stress that efficiently channels the gravitational energy into the mechanical outflow energy. Determining the precise physical mechanism that connects accretion to outflow is one of the key issues in star formation since it regulates the accretion rate and the evolution during the pre-main sequence phase. Recent UV observations have confirmed the existence of a dense, hot component (Te $\approx 10^5$ K) to the outflow as well as to the accretion flow that ought to be formed very close to the star e.g. deep into the stellar gravitational well. Therefore the UV range is instrumental to study the physics of accretion and outflow in low mass pre-main sequence stars. Moreover, it seems likely that a similar physics applies to other relativistic astrophysical systems such as QSOs, AGNs or microquasars.

Through the imaging programme, the WSO will also be able to study star formation in the nearby extragalactic Universe. The collateral survey (see section 4.7) will reveal many UV galaxies. In particular emission from newly formed stars of type earlier than A, observed at wavelengths below 160nm, will allow study of the co-moving star formation density in the low-Z Universe.

### 4.5 Active galactic nuclei

WSO/UV can considerably improve our understanding of accretion processes onto supermassive black holes in active galaxies, as their main energy output is in the far UV/extreme UV domain, and rich emission lines and occasionally absorption lines associated with surrounding gas are visible in the far UV band. The WSO will be able to address a number of problems.

WSO will be able to identify the energy source and probe structures in the UV emission region in weak active galaxies, such as LINER, Seyfert 2, H II region and starburst galaxies, through UV spectral diagnosis and UV image of the center nucleus. UV spectra can also be use to probe the structure of active galaxies and quasars through monitoring spectral variability. Multi-frequency monitoring is a very powerful tool for understanding the structure of AGN, such as the structure BLR, the size of continuum emission region as already illustrated by IUE observations. With much better S/N ratio, and spectral resolution, the WSO can do a real reverberation mapping of BLR, and measure small time lag in the continuum before different bands. The mass outflows in quasars and star-forming galaxies can be traced through studying UV absorption lines. The intrinsic absorption lines in high redshift quasars are accessible to the ground telescopes, but at low redshift, they can be studied only in the UV. With the recent discovery of special properties of BAL QSOs in the X-ray, an efficient selection of candidates is possible.

### 4.6 Structure of the Universe

WSO can be use to explore the structure of the universe through detecting the absorption feature of intervening material. As quasars are the most luminous UV emitter in the universe, the absorption lines imprint in the observed quasar spectrum when the light passes through the gas in the universe, which trace the distribution of baryon. WSO/UV can improve our understanding of this problem in several aspects: (1) detection of such absorption line systems at low redshift (z<1.5) in a statistically significant number of objects, allowing a detailed comparison with high redshift objects, such as column density and metallicity. (2) measuring the OVI and other highly ionized absorption lines, as recently observed toward some clusters of galaxies, to detect the highly ionized baryon content at low redshift. (3) measuring HeII Gunn-Peterson effect in high Z quasars to put constraints on the ionization effect and the origin of EUV background radiation.

**4.7 Collateral imaging surveys**

While WSO will observe a spectroscopic target, it is expected that all five imagers will operate in parallel mode. The particular targets viewed by any of the imagers will be known in advance, if proper observation planning will take place. Thus, it is likely that the imaging science team will be able to optimize the exposure times and the filters to be used for each pointing (to minimize confusion, for example). Parallel surveys are an important source of astronomical information because they represent unbiased selections of astrophysical objects. For studies of the stellar population of hot, evolved stars in the Milky Way galaxy such observations will be unique, because they will extract a UV-selected, unbiased sample that will reach to large distances from the galactic disk. When the WSO will point at high galactic latitudes, toward regions with reduced interstellar extinction, one would obtain a similarly unbiased sample of extragalactic objects. These would form a unique and deep sample of galaxies, from which one could evaluate properly the co-moving density of star formation in conjunction of ground-based redshifts), or the UV morphology of nearby (z<0.2) galaxies, etc. Using a number of UV bands combined with ground-based optical photometry it may be possible also to provide an unbiased sample of UV-selected AGNs, using photometric separation and redshift estimation. Given the plethora of possibly interesting (but serendipitous) objects surveyed in this mode, the parallel survey mode has the potential of finding some unique phenomena.

**4.8 Unpredictable phenomena**

The nature of Astrophysics is that many breakthroughs in the field are the consequence of unforeseen or unpredicted occurrences. We mention here discoveries of comets and their behaviour during their passage near the Sun, Novae, Supernovae, Gamma-ray Bursters, OVV's and others, are a strong components in the science addressed by the WSO/UV. It is specifically the rapid response capability required in the mission, which will present new major challenges to the scientists.

## 5. CONCLUSIONS

We have presented a description of the proposed World Space Observatory UV telescope and its instrument complement. This mission provides UV spectroscopy and imaging that is a significant improvement on that now available with HST and will be a timely replacement, taking UV astronomy forward into the second decade of the 21$^{st}$ century to build on the legacy of HST and the forthcoming GALEX UV sky survey. Since the mission concept does not rely on radical new technological developments, it can be achieved in a timely and cost-effective way filling a time-gap in access to the UV band that will inevitably emerge in advance of the development of a large UV mission matching the scope of NGST.

The approach to the development and operation of the WSO is quite different to any previous space project. However, it offers unique opportunities for true worldwide participation in space research and a means of developing education and research systems for developing countries in particular.

## ACKNOWLEDGEMENTS

MAB acknowledges support received from the International Astronomical Union and the Particle Physics and Astronomy Research Council in the UK.

* mab@star.le.ac.uk; phone +44 116 252 3492; fax +44 116 252 3311; http://www.star.le.ac.uk/~mab; Department of Physics and Astronomy, University of Leicester, University Road, Leicester, LE1 7RH, UK.